# FedBayes: A Zero-Trust Federated Learning Aggregation to Defend Against Adversarial Attacks

Marc Vucovich[a]  Devin Quinn[a]  Kevin Choi[a]  Christopher Redino[a]  Abdul Rahman[a]  Edward Bowen[a]

[a]Deloitte & Touche LLP

*Abstract*—Federated learning has created a decentralized method to train a machine learning model without needing direct access to client data. The main goal of a federated learning architecture is to protect the privacy of each client while still contributing to the training of the global model. However, the main advantage of privacy in federated learning is also the easiest aspect to exploit. Without being able to see the clients' data, it is difficult to determine the quality of the data. By utilizing data poisoning methods, such as backdoor or label-flipping attacks, or by sending manipulated information about their data back to the server, malicious clients are able to corrupt the global model and degrade performance across all clients within a federation. Our novel aggregation method, *FedBayes*, mitigates the effect of a malicious client by calculating the probabilities of a client's model weights given to the prior model's weights using Bayesian statistics. Our results show that this approach negates the effects of malicious clients and protects the overall federation.

## I. Introduction

Federated Learning (FL) has become an emerging field within artificial intelligence (AI) and machine learning (ML) due to the ability to aggregate insights from multiple clients without having direct access to the client's data. FL works by initializing a server that orchestrates the training of the clients participating in the federation. Once the needed number of clients connect to the server, they train their models on their local data. The clients' new model weights are then sent back to the server where they are aggregated together to form the new global model. This process is repeated for a pre-determined number of rounds [1]. The main aggregation methods used today are variations of *FedAvg* [1], which averages each client's weights together to form the global model. These aggregation methods, however, are particularly vulnerable to adversarial attacks such as backdoor attacks, label flipping, and weight attacks. Our paper proposes a new FL aggregation method called *FedBayes*, which calculates the probability of receiving a client's weights given the prior global model's weights such that the received weights are not detrimental to the federation.

*FedBayes* is a novel aggregation method created to reduce the effect of corrupted clients within the federation while still learning from the benign clients. For each layer in the global model, we calculate the normalized distribution using the mean and standard deviation of the model's weights. Using a cumulative distribution function, we calculate the probability of each weight received by the clients given the cumulative distribution function of the global model's weights. Since poisoned weights are expected to be outside of the normal distribution of the clean global model, and a weight with a low probability has less influence than a weight with a high probability, the effects of malicious clients upon aggregation are reduced on the updated model.

This paper is structured as follows: Section 2 provides a background of federated learning and adversarial attacks. Section 3 highlights related works around other trust-based and probability-based methods that have helped us create *FedBayes* as well as their potential advantages and disadvantages. Section 4 discusses the datasets used for the experiments. Section 5 describes our methodology for *FedBayes*, while Section 6 explains our experimental design and Section 7 discusses our results. Finally, Section 8 ends with our takeaways and thoughts for continued research.

## II. Background

### A. Federated Learning

Typically, an ML model is trained with all the data in one location. This is referred to as centralized training since all of the data is in one central location. On the other hand, FL is a decentralized training technique that allows a global model to learn from data on remote servers or locations [2]. In an FL system, multiple clients train a model using their local data and send their trained model's weights back to the server [1]. Once the server receives the weights, the weights are typically averaged to create a global model [1]. Once the clients complete this process over a set number of iterations or rounds, the training is complete, and each client has a copy of the global model that is an aggregation of all models within the federation [1]. FL can help protect the privacy of client data, but security risks are still a big concern [2]. The method in this paper helps mitigate these concerns.

### B. Adversarial Attacks

Adversarial attacks refer to the class of methods used to control a model's prediction through data manipulation, some via the incorporation of adversarial examples [3]. Adversaries aim to exploit victim models for a malicious goal where a change in classification would be to their advantage. This not only raises concerns around the reliability of model predictions, but also brings serious consequences for security-critical tasks. For example, Chen et al. [4] show adversarial attacks can be utilized to bypass a facial recognition system to gain unauthorized access to a security system.

Adversarial attacks can be targeted, where an adversary attempts to trick a victim model into misclassifying an input to be a hand-picked class, or untargeted, where an adversary only aims to degrade model predictions overall [5]. Attacks can further be classified based on the threat model of the victim. White-Box Attacks refer to the scenario where a malicious actor has full knowledge or access to the victim model and training data set. Conversely, in Black-Box Attacks an adversary has no direct knowledge of the victim model or training data. In Gray-Box scenarios, adversaries only have limited knowledge of the victim model and/or training data [5].

### C. Types of Adversarial Attacks

There are several different types of adversarial attacks depending on how the input data is manipulated, the goal of the adversary, and the environment the attack is conducted in. Since federated learning involves the simultaneous training of multiple models, we focus on data poisoning attacks.

Data poisoning involves the adversarial contamination of the training data and/or pre-trained weights to manipulate a model during the training phase. There are several different types of data poisoning attacks; the ones focused on in this paper are label-flipping attacks and backdoor (trojan) attacks. In a label-flipping attack, an adversary simply changes the labels of a class in the training set, either targeted or untargeted, to degrade prediction performance on this class [6]. In a backdoor attack, an adversary embeds a specific trigger into the data samples of a certain class and changes the labels of these manipulated samples to a target label. The motivation is to train a backdoor into the model where an association is made between the trigger and the target label such that new inputs with the trigger are predicted to be the target class during inference [7]. Adversaries can train backdoor triggers into pre-trained models and have their effects persist through transfer learning, creating serious security concerns for researchers looking to leverage open-source pre-trained models [8].

Poisoning attacks conducted specifically in a collaborative learning environment, such as a FL system, are known as Byzantine Attacks. These types of attacks involve one or a few malicious client models within a federation corrupting their own models and spreading poisoned weights during the global aggregation step, thereby infecting all client models within the system [9]. Byzantine attacks may also involve exploiting FL specific steps to attack clients, such as conducting a weight attack [10]. In this type of attack, an adversary exploits the update aggregation step where the weight assigned to each update depends on the size of the local training set reported by the client. Since the central server has no authority to check the validity of these reported sizes (due to client privacy reasons), a malicious client can lie about its data set size to gain a high weight, resulting in the poisoned weights having the strongest influence over the global model.

## III. LITERATURE REVIEW

Although FL is a decentralized method to train an ML model without accessing client data, there are still security risks and aggregation difficulties that researchers have been looking to diminish [2].

### A. Performance-based Trust Methods

Performance-based methods attempt to defend against adversarial attacks by evaluating over a clean data set provided by the central server, where poorly performing updates will have little to no influence over the global model. Xie et al. propose Zeno [11], a method utilizing a distributed Stochastic Gradient Descent (SGD) approach where during each iteration, a client pulls the latest central server model and estimates the gradients using their locally sampled training data. A stochastic zero-order oracle is used to compute a score, ranking how trustworthy each client update is in that iteration and averaging over the clients with the highest score.

Cao et al. propose FLTrust [12], which bootstraps trust by utilizing a rectified linear unit (ReLU) clipped cosine similarity between local and server updates calculated on a clean data set to assign trust scores to clients. Local model updates have a lower trust score if its direction and magnitude is inconsistent to that of the server model update. The central server then normalizes the local model updates weighted by their trust scores as a global model update.

These methods can be difficult to implement in practice as they rely on having an adequate clean data set. If the evaluation data is not sufficiently generalized, which can be difficult to achieve across different clients, benign client updates may differ from the server updates on the insufficient data set, hindering these clients from contributing to the system. Conversely, an insufficient evaluation set that is too generalized may result in malicious updates not greatly differing from the server updates, allowing poisoned weights to reach the global model. Additionally, methods that assign scores and weight the updates can incorrectly assign a high weight to a missed malicious client, giving larger influence to a poisoned model and inadvertently result in a weight attack on the system.

### B. Probability-based Aggregation Methods

Liu et al. [13] provide a method to aggregate weights for Bayesian models using heterogeneous data. To do this, they follow a two-part process. On the server side, they multiply the posteriors of the Gaussian normal distribution for each client's weights to create a global Gaussian normal distribution. This helps reduce the aggregation errors induced by heterogeneous data [13]. On the client side, the authors propose a prior iteration (PI) strategy that treats the global posterior probabilistic parameters distributed from the server as priors. With PI, they derive a prior loss from the prior distribution for local training. This is meant to help prevent clients from drifting away from the global posterior while training [13]. This approach has many benefits including the ability to handle heterogeneous data. The biggest takeaway from this paper is the use of Bayesian probability to find a compromise between heterogeneous clients. However, in FL, it is imperative to have a zero-trust approach while training since it is easy for a client or a hacker with access to a client to alter the results sent back to the server. Although it is unlikely that a client will manipulate the results being sent back to the server, relying on using the PI on the client side to influence the local training presents a risk of the client corrupting the global model. With that in mind, we set out to build upon the Bayesian aggregation method by removing the reliance on trustworthy clients.

Nguyen et al. propose FedProb [14], a method that calculates the multivariate normal distribution for each client by using the mean and covariance of each client's local Non-IID (independent and identically distributed) dataset. This information is sent to the server where the distributions of the datasets are averaged to create a global distribution. The authors then calculate the distance between each client distribution and the global distribution to determine the amount of influence each client has on the new global model [14]. Similar to the Bayesian approach above, the method is able to achieve strong performance on Non-IID data, but it presents a few potential security risks. One risk would be that a hacker would be able to gain insights about client data by intercepting the provided weights, as well as the mean and covariance of the data when they are sent to the server.

We set out to to create a method which calculates the similarity between clients without relying on them to be trustworthy. Both of these probability based approaches are comparable to our approach and influenced the aggregation method presented in this paper, which helped lead to a secure FL architecture that is able to prevent malicious acting clients from corrupting the overall federation.

## IV. Datasets

In the experiments documented below, the Modified National Institute of Standards and Technology (MNIST) and Canadian Institute for Advanced Research (CIFAR-10) datasets were used with additional customization to replicate attack scenarios.

The MNIST dataset contains 60,000 training examples and 10,000 testing examples of low resolution hand-written digits from 0-9 [15]. The CIFAR-10 dataset contains a collection of 60,000 32x32 pixel color images with 10 different classes [16]. For the experiments in this paper, the data was split into 9 subsets; one dedicated to the initial pre-trained model and the other eight to be used as our clients. The subsets were roughly around the same size to clearly demonstrate the effects of a weight attack in our experiments. The classes were also balanced and equally represented across each set, making the data IID across the clients.

In a backdoor attack, a portion of the data is manipulated to create a 'marker' that trains the model to predict the wrong value when the 'marker' appears during inference. For the backdoor experiments, 70% of the data in one subset (client 1) had a cross shape embedded into in the upper left corner of the image. In the MNIST experiments, the manipulated images had their labels set to '2'. In the CIFAR-10 experiments, the manipulated images had their labels set to 'bird'. Additionally, the cross trigger was added to 50% of the examples in the respective test sets but the labels remained as the original labels. This allowed us to to test the effects of the backdoor signal on the global model. Given the backdoor marker was added to 50% of the test data, a corrupted model would be expected to achieve peak performance of approximately 50% accuracy. A sample of the data in the poisoned set with some backdoored examples is illustrated in Fig. 1.

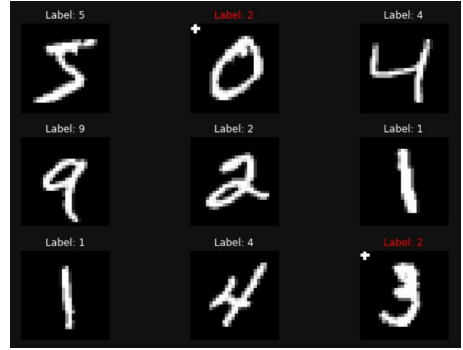

**Fig. 1:** Subset of MNIST training data with random backdoor samples

In a label-flipping attack, a portion of the labels for the input data are changed to be an arbitrary incorrect value. Although not as powerful as backdoor attacks in terms of prediction manipulation, this attack is still able to degrade model performance and affect the global model. For the label-flipping experiments, 85% of client 1's data had the labels switched. In the MNIST experiments, the poisoned examples had their labels set to '2'. In the CIFAR-10 experiments, the poisoned examples had their labels set to 'bird'. These labels were switched for examples that don't already have these ground truth labels so that 85% of the data was actually manipulated.

## V. Methodology

There have been a variety of FL aggregation techniques created in recent years, but most approaches aggregate data with the assumption that each client will be acting to improve the global model. *FedBayes* is an aggregation algorithm that not only learns as well as the standard FL aggregation methods but also mitigates the effects of malicious clients trying to corrupt the global model. In some attack scenarios, *FedBayes* even allows the federation to continue learning from benign clients while ignoring the malicious clients.

*FedBayes* is intended to be used with a pre-trained model that the server will send to the clients as the initial parameters. These initial parameters will serve as the prior in the probability calculations. Once the clients have the initial parameters, the clients will continue to train the model using their own data. Each client will then send the updated parameters back to the server. It is important to note that the client does not need to send any additional information to the server besides the model parameters.

On the server side, the algorithm loops through each layer of the initial parameters, which will be referred to as the prior parameters, and calculate the mean and standard deviation for that layer's weights. The prior parameter's mean and standard deviation will be referred to as the global mean, $\mu_{global}$ and global standard deviation, $\sigma_{global}$. Using the global mean and global standard deviation of the prior parameters, the global normal distribution is calculated and the global cumulative distribution function (CDF) is created. With the global CDF, the cumulative probability of receiving a client's parameters is calculated, given the prior parameters. To do this, the CDF of the clients' weights are subtracted from the CDF of the prior parameters.

$$P(\mathbf{Cw}|\mathbf{Gw}) = \Phi(\mathbf{Gw}, \mu_{global}, \sigma_{global}) - \Phi(\mathbf{Cw}, \mu_{global}, \sigma_{global}) \quad (1)$$

If a client's weights are similar to the global weights, the resulting value will be closer to zero. Therefore, we subtract the value from 1. This final value is the probability of receiving a clients weights given the prior parameters.

$$\mathbf{Pn} = 1 - P(\mathbf{Cw}|\mathbf{Gw}) \quad (2)$$

However, we found the difference between a malicious client's weights and a benign client's weights to be on a very small scale. To counter this, we added penalty factor of 100 for each percent the probability is away from 1. For example, if the value given from $P(\mathbf{Cw}|\mathbf{Gw})$ is 0.003, the value is adjusted to become 0.3. This value is then subtracted from 1 to give a final probability of 0.7. The final formula is shown below.

$$\mathbf{Pn} = 1 - (100 * (P(\mathbf{Cw}|\mathbf{Gw}))) \quad (3)$$

Finally, to aggregate the weights we multiply the clients weights by the adjusted probability, and divide by the sum of each clients adjusted probability. This approach allows the server to ignore the feedback received from malicious clients. This is shown in the equation below where n is the number of total clients.

$$Weights = \frac{\sum_{n=1}^{n} (\mathbf{Cwn} * Pn)}{\sum_{n=1}^{n} (\mathbf{Pn})} \quad (4)$$

## VI. EXPERIMENTAL DESIGN

The experiments in this paper were set up to accomplish three objectives: show *FedBayes* is able to learn as well as the standard FL aggregation methods with clean data, show *FedBayes* is able to mitigate the effects of a backdoor attack better than the standard FL aggregations, and show *FedBayes* is able to mitigate the effects of a label-flipping attack better than the standard FL aggregations.

**Algorithm 1** FedBayes

**Input:** Client weights ($Cw$), Prior / Global weights ($Gw$)
**Output**: Final Weights ($W$)

1: $AdjustedWeights(Aw) \leftarrow []$
2: $Probabilities(P) \leftarrow []$

3: **for** layer in $Gw$ **do**
4:     $\sigma_{global}$ = std of prior parameters
5:     $\mu_{global}$ = mean of prior parameters
6:     adjusted layer weights ($Lw$) $\leftarrow []$
7:     layer probabilities ($Lp$) $\leftarrow []$

8:     **for** client in Federation **do**
9:        $Pn \leftarrow 1 - (100 \times (P(Cw \mid Gw)))$
10:        Append ($Cw \times Pn$) to $Lw$
11:        Append ($Pn$) to $Lp$
12:     **end for**

13:     Append $Lw$ to $Aw$
14:     Append $Lp$ to $P$
15: **end for**

16: $W \leftarrow \frac{\sum_{n=1}^{n} Aw_n}{\sum_{n=1}^{n} P_n}$

17: **return** $W$

We compare the adversarial robustness of *FedBayes* to *FedAvg* [1], *FedAdam* [17], *FedAdagrad* [17], and *FedYogi* [17]. We simulate a FL environment with these algorithms using the Flower Framework [18].

For experimentation, the training data set was split into 9 subsets while the test set was kept separate. The first subset was used to train an initial pre-trained model in the FL setup. The pre-trained model is also used as a benchmark to show whether the model has learned or been corrupted. The MNIST experiments used a simple Convolutional Neural Network (CNN) model architecture, and the pre-trained model was trained to be approximately 80% accurate on the test data. The CIFAR-10 experiments used a Residual Neural Network (ResNet) model architecture, and the pre-trained model was trained to be approximately 60% accurate on the test data.

The remaining 8 subsets of data were used as the 8 clients in the FL set-up. Each client had slightly varying amounts of examples of each class in the dataset, but they all contained examples of every class. The FL training consisted of 100 total aggregation rounds, with each round having 5 local model epochs. Additionally, we adjusted the aggregation hyper-parameters for *FedAdam*, *FedAdagrad*, *FedYogi*, such that they had optimal performance with the clean data. After establishing a strong baseline performance for each, we carry the respective hyper-parameters of each algorithm and experiment using the poisoned data sets.

For these experiments, we limit the number of poisoned

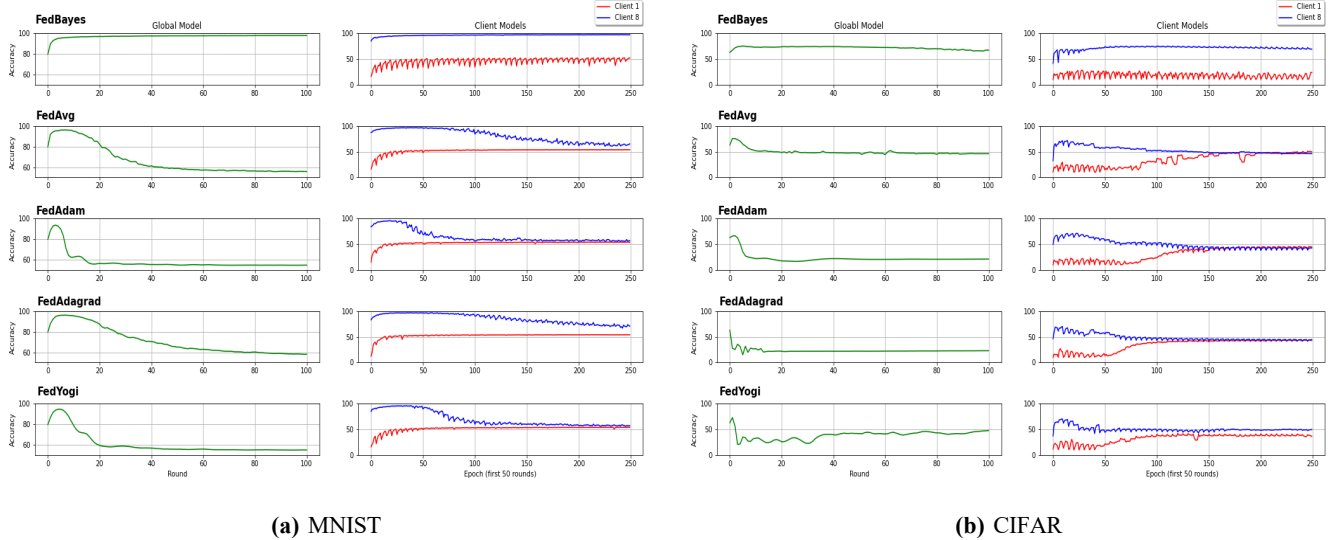

**(a)** MNIST      **(b)** CIFAR

**Fig. 2:** Backdoor attack (no weight) strategy performance comparison over 100 rounds of training, with 5 epochs per round for each client model. The global (aggregated) model performance is shown on the left sides of 2a and 2b while individual model performance of client 1 (malicious) and client 8 (benign) over the first 50 rounds is shown on the right sides. The other 6 benign clients show similar performance to client 8, but are not shown here.

data sets to 1 of 8 clients. However, we should note that we've found *FedBayes* fail to stop the global adversarial effects when too many malicious clients are introduced. We conducted a separate experiment to test the limits of the algorithm. After increasing our total number of clients and varying the number of poisoned data sets, we found *FedBayes* begin to fail when more than 35% of the clients are malicious. Although this is a point of failure for the algorithm, we find this to not be a practical issue as it is unlikely you would find that many malicious clients in a realistic federation.

The experiments are laid out in three sections:
1. Baseline Performance: Compares the overall learning ability of *FedBayes* versus *FedAvg*, *FedAdam*, *FedAdagrad*, and *FedYogi* when trained using clean data.
2. Backdoor: Compares how the above aggregation methods perform when a client is performing a backdoor attack.
3. Label-Flipping: Compares how the above aggregation methods perform when a client is performing a label-flipping attack.

## VII. RESULTS

Please note that each experiment shown in the results started with a pre-trained model. Again, the MNIST pre-trained model was approximately 80% accurate on the test data and the CIFAR-10 pre-trained model was approximately 60% accurate on the test dataset. These accuracies will serve as the benchmark performance to show when the FL model has learned properly or was corrupted from the malicious client data.

### A. Baseline Performance

The first set of experiments show the performance of *FedBayes* compared to standard FL aggregation methods using clean data. The first baseline experiment used MNIST data while the second baseline experiment used CIFAR-10 data; these results are shown in Fig. A.1.

In A.1a and A.1b, the global accuracy (shown on the left sides of the subfigures), increases as expected for all aggregation methods. Additionally, the accuracies plotted in the charts on the right show that client 1 and client 8 learn throughout each epoch. These results demonstrate that *FedBayes* is able to learn as well as *FedAvg*, *FedAdam*, *FedAdagrad* and *FedYogi* in a normal collaborative learning environment and can be used in place of these aggregation methods.

### B. Backdoor Attacks

This experiment was run twice for each dataset. The first set of experiments did not use a weight attack; this means that each client reported the correct amount of data used in training back to the server and all clients have an equal weighting. The results of this experiment are shown in Fig. 2. Analyzing the global model accuracies on the left sides of 2a and 2b, it is apparent that *FedBayes* performs better than the other standard FL aggregation methods. In the case of the MNIST experiments, which is a simpler dataset, the global model is clearly able to learn even with the malicious client, quickly converging to 98% accuracy on the test set and keeping steady performance through the 100 rounds. Even with CIFAR-10, the final accuracy of the global model was 67% accurate, which is an improvement from the initial pre-trained model.

The charts on the right of these subfigures show the individual model performance of clients 1 (red, malicious) and 8

(blue, benign) during the first 50 rounds of training. These results emphasize *FedBayes'* ability to protect the benign clients in the federation from the malicious client, while the other algorithms show the benign models' accuracy eventually drop and converge to malicious performance.

The second set of experiments combine a backdoor attack with a weight attack. In these experiments, client 1, the malicious client, reported using twice the amount of data in order to have more influence on the global model. We believe the combination of a backdoor attack with weighting has the most profound adversarial effect of these experiments, and we highlight the ability of *FedBayes* to negate these effects in Fig. A.2. We see similar performance to the previous experiment, but the weighting causes a quicker drop in performance for the other algorithms besides *FedBayes*, which is entirely unaffected by the weighting.

*C. Label Flipping*

This experiment was run twice for each dataset. Similar to the backdoor experiments, the first set did not use a weight attack while the second combined label-flipping with weighting. This time, we use a 3x weight attack such that malicious client 1 is reporting three times the amount of data it actually has.

The results from the no weight label-flip attack is shown in Fig. A.3. Although all of the aggregation methods are able to handle the label-flipping attack fairly well, we see some unstable performance in the other methods. This unstable performance is exaggerated in the 3x weighted attack shown in Fig. A.4, which shows *FedBayes* has a much easier time learning and negating the adversarial effects than the other strategies. Here, we see the other aggregations have a lot more oscillations due to the malicious client impacting the global model updates while *FedBayes* achieves the strongest performance while maintaining stability throughout the 100 rounds.

These results show that while introducing a label-flipping adversary into a federation doesn't force the benign clients' accuracy to drop and converge to malicious levels of performance as it does in the backdoor scenario, the label-flipped data can still disrupt the process and cause unstable training, an effect *FedBayes* largely negates.

## VIII. CONCLUSION

The results from the experiments show the ability of *FedBayes* to protect all of the clients participating in the federation from a malicious acting client. Additionally, *FedBayes* is able to learn as well as the standard FL aggregation methods currently used, providing clients with a more secure alternative to conventional methods. *FedBayes* becomes particularly useful in scenarios where clients are working together to create a global model to protect valuable information. Imagine a group of companies are working together to build an intrusion detection system (IDS) that detects incoming attacks. In this scenario, if a client or group of clients label known attacks as benign and train their model with the data, the global model may become vulnerable to these attacks after the FL training. With *FedBayes*, the clients in the federation are better protected from the corrupted data. Overall, *FedBayes* is a simple aggregation technique that protects the global model from adversarial attacks while still learning from benign clients.

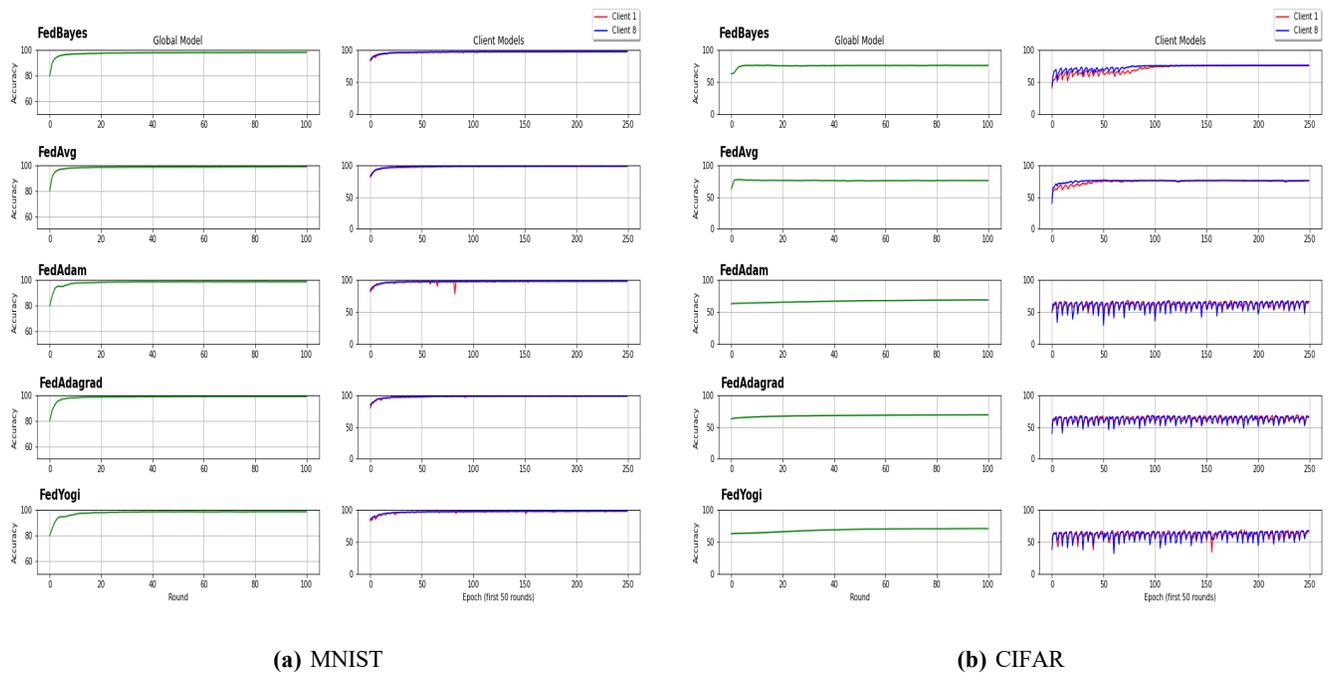

(a) MNIST

(b) CIFAR

Fig. A.1: Baseline comparison

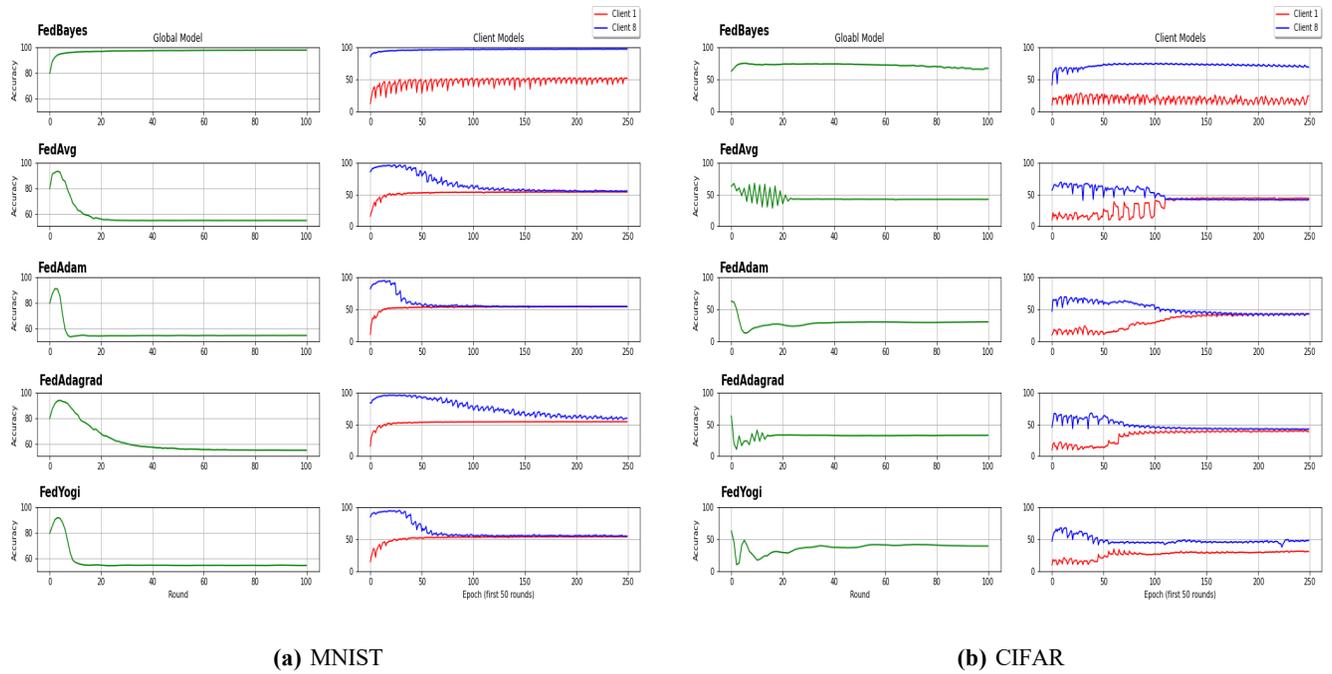

(a) MNIST

(b) CIFAR

Fig. A.2: Backdoor attack (2x weight) comparison

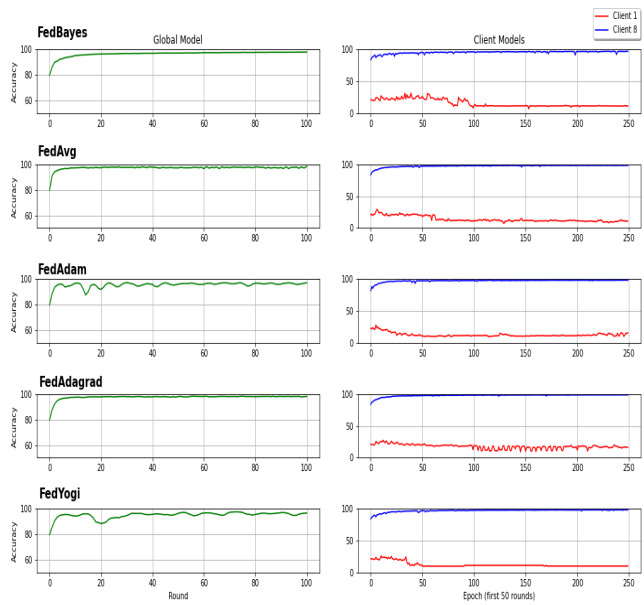
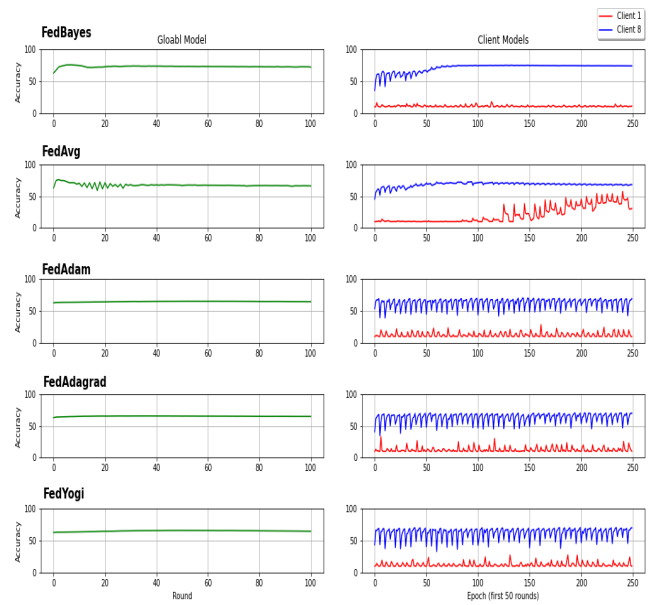

**(a)** MNIST

**(b)** CIFAR

**Fig. A.3:** Label-flip attack (no weight) comparison

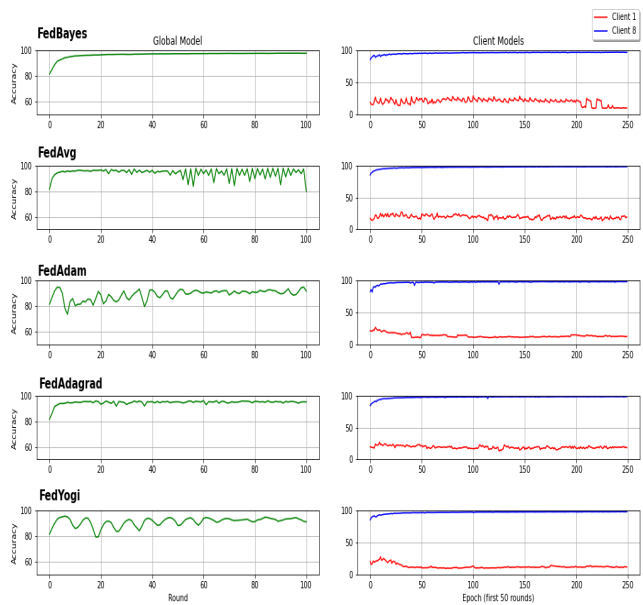
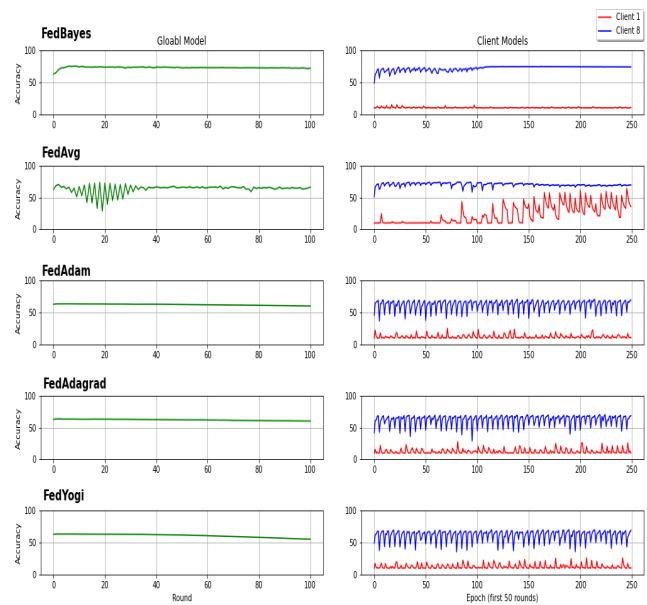

**(a)** MNIST

**(b)** CIFAR

**Fig. A.4:** Label-flip attack (3x weight) comparison